\documentclass[twocolumn,twocolappendix]{aastex631}

\usepackage{graphicx}	% Including figure files
\usepackage{amsmath}	% Advanced maths commands
\usepackage{subfigure}
\usepackage{natbib}
\usepackage{color}
\usepackage{tabularx}
\usepackage{longtable}
\usepackage{bm}
\usepackage{threeparttable}
\usepackage{enumerate}
\usepackage[normalem]{ulem}
\usepackage{verbatim}
\usepackage[T1]{fontenc}

\newcommand{\code}[1]{{\texttt{#1}}}
\newcommand{\tess}{{\it TESS}}

\newcommand{\tar}{{TOI-5375}}

\begin{document}

%\title{Spin-Orbit Alignment for TOI-4201b: an M Dwarf with a Hot Jupiter}
\title{An Aligned Very-Low-Mass Star Orbiting an M dwarf and Obliquity Patterns Across Giant Planets, Brown Dwarfs, and Binary Stars}

\correspondingauthor{Tianjun Gan}
\email{tianjungan@gmail.com}

\author[0000-0002-4503-9705]{Tianjun~Gan}
\affil{Department of Astronomy, Westlake University, Hangzhou 310030, Zhejiang Province, China}

\author[0009-0005-6135-6769]{Alexandrine L'Heureux}
\affil{Institut Trottier de recherche sur les exoplan\`etes, D\'epartement de Physique, Universit\'e de Montr\'eal, Montr\'eal, QC H3C 3J7, Canada}
%\affil{Universit\'e de Montr\'eal, D\'epartement de Physique, IREX, Montr\'eal, QC H3C 3J7, Canada}

%SPIRou member
\author[0000-0003-3506-5667]{\'Etienne Artigau}% RV
\affil{Institut Trottier de recherche sur les exoplan\`etes, D\'epartement de Physique, Universit\'e de Montr\'eal, Montr\'eal, QC H3C 3J7, Canada}
\affiliation{Observatoire du Mont-M\'egantic, Universit\'e de Montr\'eal, Montr\'eal, QC H3C 3J7, Canada}

\author[0000-0001-9291-5555]{Charles Cadieux}% RV
\affil{Institut Trottier de recherche sur les exoplan\`etes, D\'epartement de Physique, Universit\'e de Montr\'eal, Montr\'eal, QC H3C 3J7, Canada}
\affil{Observatoire de Gen\`eve, D\'epartement d’Astronomie, Universit\'e de Gen\`eve, Chemin Pegasi 51, 1290 Versoix, Switzerland}

\author[0000-0001-5485-4675]{Ren\'e Doyon}% RV
\affil{Institut Trottier de recherche sur les exoplan\`etes, D\'epartement de Physique, Universit\'e de Montr\'eal, Montr\'eal, QC H3C 3J7, Canada}
\affiliation{Observatoire du Mont-M\'egantic, Universit\'e de Montr\'eal, Montr\'eal, QC H3C 3J7, Canada}

\author[0000-0003-4166-4121]{Neil J. Cook}% RV
\affil{Institut Trottier de recherche sur les exoplan\`etes, D\'epartement de Physique, Universit\'e de Montr\'eal, Montr\'eal, QC H3C 3J7, Canada}

\author[0000-0001-8317-2788]{Shude Mao}
\affil{Department of Astronomy, Westlake University, Hangzhou 310030, Zhejiang Province, China}



\begin{abstract}
Stellar obliquity serves as a key diagnostic for tracing the dynamical evolution of bound systems—from giant planets and brown dwarfs to stellar binaries—revealing whether these diverse populations share analogous histories. Here, we report the first obliquity measurement for a double M dwarf system, determined via the Rossiter–McLaughlin effect. The spin axis of the primary star, TOI-5375 ($M_\ast=0.62\pm0.02\,M_\odot$), is well aligned with the orbit of its low-mass stellar companion ($M_c=84.8\pm1.5\, M_J$, $P=1.72$\,days) with a projected obliquity of $\lambda=-13.5_{-13.8}^{+12.4}\,^{\circ}$ and a true 3D obliquity of $\psi=37.5_{-13.4}^{+10.6}\,^{\circ}$. The result indicates that the system either formed with a primordially aligned configuration or has undergone tidal realignment. We further investigate obliquity patterns across giant planets, brown dwarfs and binary stars. It turns out that a few obliquity trends observed in giant planets also tentatively exhibit in the latter two higher-mass populations: 1) well-aligned orbits are preferentially found around cooler host stars ($T_{\rm eff}\leq 6250\,K$); 2) wide-orbit ($a/R_\ast\geq 10$) companions are predominantly aligned; 3) no significant correlation shows up between obliquity and orbital eccentricity in any of the companion classes. By modeling $|\lambda|$ with a two-component Gaussian distribution, we find that the low-$|\lambda|$ components of binary stars and brown dwarfs are more concentrated near zero than giant planets while the high-$|\lambda|$ components of brown dwarfs and binaries remain unclear due to the small sample size. 
\end{abstract}



\section{Introduction} \label{sec:intro}

Binary stars are ubiquitous in the Milky Way \citep{Duquennoy1991}. Approximately half of nearby Sun-like stars have at least one stellar companion \citep{Raghavan2010}, and the multiplicity rate decreases to about 20\% when moving to low-mass M stars \citep{Ward-Duong2015,Clark2024,Cifuentes2025}. Regarding the binary star formation channels \citep[see][for reviews]{Tohline2002,Duchene2013,Offner2023}, several mechanisms have been postulated, including core fragmentation of turbulent protostellar cores \citep{Wurster2018,Lee2019,Kuruwita2023}, disc fragmentation within massive, cold and gravitationally unstable circumstellar discs \citep{Adams1989,Kratter2016,Tokovinin2020} as well as dynamical capture from unbound flyby encounters \citep{Parker2014,Murillo2016}. 

While the aforementioned pathways are effective at wide separations ($\sim 100$~AU), they struggle to explain the formation of tight-orbit ($\sim 0.05$~AU) binary stars \citep{Moe2018}, which have been frequently detected in both ground- and space-based surveys \citep[e.g.,][]{Norton2011,Soszynski2016,Prsa2022}. Consequently, the short-period stellar binary population is thought to originate from subsequent evolution mechanisms that push them inward after formation via, for example, interactions with a third body \citep{Eggleton2006,Fabrycky2007,Naoz2014} or disc-driven migration through dynamical friction with gas \citep{Lee2019,Tokovinin2020}. These channels are similar to those invoked for giant planet systems \citep{Dawson2018}. 

The different formation and evolution processes of binary stars may imprint distinct signatures on the orbital eccentricity and stellar obliquity (i.e., the angle between the spin axis of the host star and the orbital angular momentum vector of the companion). These distributions can therefore serve as diagnostic probes of their origins. Based on a sample of transiting stellar companions with companion masses  $80\ M_J \leq M_c \leq 150\ M_J$ and orbital periods exceeding 10 days, \cite{Gan2025} found that these low-mass-ratio binaries tend to present an eccentricity peak at about 0.3. Such a feature is similar to that seen in wide-orbit massive binaries \citep{Duquennoy1991,Wu2025}, indicating a similar evolution history. In terms of the stellar obliquities of double star systems, however, only a few results are available so far \citep[e.g.,][]{Albrecht2013,Triaud2013,Gill2019,Kunovac2020,Marcussen2022,Wells2025,Spejcher2025,Spejcher2025b}, determined based on the Rossiter-Mclaughlin (RM) effect \citep{Rossiter1924,McLaughlin1924} by tracking the spectral line distortion during the transit. Notably, the stellar obliquities of M dwarf-M dwarf binaries have never been investigated before, despite existing studies of three M dwarf-hosted giant planets \citep{Gan2024,Weisserman2025} and two brown dwarfs \citep{Brady2025,Doyle2025}. Although several obliquity trends have been reported for giant planets, such as 1) hot Jupiters around hot stars tend to be misaligned \citep[e.g.,][]{Winn2010,Albrecht2012,Wang2026}; 2) warm Jupiters mostly have aligned orbits \citep[e.g.,][]{Rice2022warmjupiter,Wang2024}; and 3) high-mass-ratio giant planets are more aligned \citep[e.g.,][]{Hebrard2011_HATP6,Rusznak2025}, it remains unclear if such patterns also appear in the brown dwarf and binary star categories. 

Here, we report the first RM effect measurement for an M dwarf binary. The target, TOI-5375, was initially classified as a verified planet candidate transiting an early-M dwarf every 1.7 days \citep{Gan2023}, and later confirmed to be a very-low-mass stellar companion at the hydrogen-burning limit \citep{Lambert2023,Maldonado2023}. We then compare the obliquity behaviors of gas giants, brown dwarfs as well as binary stars, exploring their similarities and differences. The rest of this work is structured as follows. In Section~\ref{spirou_data}, we describe the spectroscopic data we obtained. Section~\ref{jointfit} presents the joint analysis we performed and the physical properties of \tar B we derived. We discuss the obliquity patterns across the three companion groups in Section~\ref{pattern}, and summarize our findings in Section~\ref{conclusion}.

%all eleven brown dwarfs with obliquity measurements: HD 33609b: \cite{Vowell2025}, CWW 89b \cite{Carmichael2025}, XO-3b \cite{Rusznak2025}, TOI-2119b \cite{Doyle2025}, LP 261-75b \cite{Brady2025}, TOI-2533b \cite{Ferreira2024}, GPX-1b \cite{Giacalone2024}, HATS-70b \cite{Zhou2019b}, WASP-30b \cite{Triaud2013}, KELT-1b \cite{Siverd2012}, CoRot-3b \cite{Triaud2009}.

\section{CFHT/SPIRou spectroscopic observations}\label{spirou_data}

We collected two transits of \tar B on UT 2025 February 10$^{\rm th}$ and UT 2025 February 17$^{\rm th}$ using SPIRou \citep[SpectroPolarim\`etre InfraROUge;][]{Donati2020} installed on the 3.6m Canada-France-Hawaii Telescope (CFHT). SPIRou is a high-resolution ($R\approx$75,000) spectrograph spanning a wavelength range from 0.98 to 2.5\,$\mu$m. The thermalized Farby-P\'erot (FP) etalon was used for wavelength calibration. Two transit observations were conducted with a 900\,s exposure time under an airmass of about 1.8 and 1.7, resulting in a signal-to-noise ratio (SNR) of 42 and 49 at Order 35 ($\sim1.75\,\mu$m), respectively. Both observations covered the 1.7\,hr full transit event along with about 1.5\,hr of baseline before the ingress and 1.5\,hr after the egress. 

The spectroscopic data were reduced using \code{APERO} version 0.7.293, the standard data reduction pipeline for SPIRou \citep{Cook2022}. The resulting telluric-corrected spectra are then used to compute the radial velocities (RV) of TOI-5375 with the line-by-line (LBL, v0.65.009) method of \citet{Artigau2022}. In this framework, the Doppler shift is independently computed for each spectral line with the \citet{Bouchy2001} formalism, allowing the identification and removal of outlying spectral features. To limit the presence of potential biases in out RV timeseries, we additionally mask atmospheric OH lines \citep[line list from][]{Rousselot2000} and oxygen fluorescence band. We obtain the final RV measurement from an error-weighted average of all the valid per-line velocities. For faint stars in particular, an important step of the LBL method is the selection of a high-SNR template spectrum against which the Doppler shift can be computed for individual spectra. This will usually be a bright star of similar spectral type as the target of interest, and with a significant amount of SPIRou observations. However, with a high spectroscopic broadening velocity of $v\sin i=16.7\pm0.9$~km s$^{-1}$ \citep{Lambert2023}, TOI-5375 is a poor match to the available SPIRou template stars, which mostly have low $v\sin i$. Instead, we combine many reference stars into a single vetted template tailored to TOI-5375 (See appendix~\ref{appendix:vetted_template}). The resulting RVs on two nights have a median uncertainty of 63 and 57\,m s$^{-1}$. 

To examine whether the RM signals were due to spot-induced rotational RV variability, we investigated the correlations between the RVs and stellar activity indicators including the chromatic RV index \citep[CRX;][]{Zechmeister2018,Artigau2022} and the differential temperature of the star \citep[dTemp;][]{Artigau2024}. For each night, we computed the Pearson correlation coefficients, and we found no significant correlations with p-values smaller than 0.05. All SPIRou RVs and stellar activity diagnostics are listed in Table~\ref{RVtable}.

\begin{figure*}[htb]
    \centering
    \includegraphics[width=0.99\linewidth]{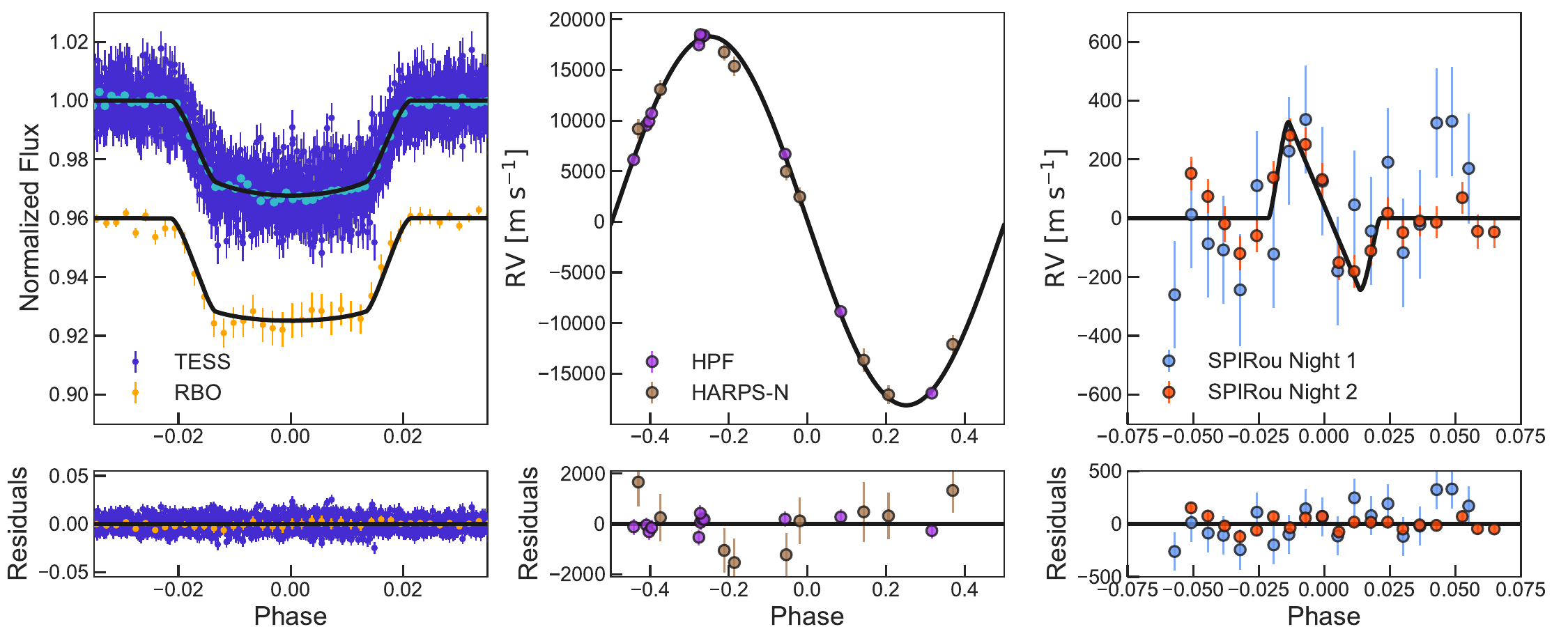}
    \caption{Left panel: The phase-folded TESS and RBO light curves. The light blue dots are the binned TESS light curve with a binning size of about 180s. Middle panel: The out-of-transit radial velocities from HPF and HARPS-N. Right panel: The two-night SPIRou RM measurements after subtracting the best Keplerian model. The plotted error bars are the quadrature sums of the uncertainties of individual measurements and the jitters. The best-fit models are shown as black curves in each plot. The residuals are presented in three bottom panels.}
    \label{fig:joint-fit}
\end{figure*}

\section{Analysis}\label{jointfit}
\subsection{Joint Fit of Transit, RV and RM Data}
We utilized the \code{Allesfitter} code \citep{allesfitter-code,Gunther2021} to jointly analyze all photometry and RVs, together with the RM data, to determine the stellar obliquity of \tar\ and refine other physical parameters. The RM model was generated based on the flux-weighted radial velocity using the \code{ellc} package \citep{Maxted2016}. The fit was carried out assuming the companion's flux does not contribute to the photometry and spectral lines, according to the companion-to-primary mass ratio \citep{Lambert2023,Maldonado2023} and luminosity-mass relation \citep{Benedict2016,Mann2019}.

%Such an assumption is reasonable as the luminosity ratio between the companion and primary star is about $10^{-3}$, based on the mass ratio $M_{c}/M_\ast$ \citep{Lambert2023,Maldonado2023} and luminosity-mass relation $L\propto M_\star^{3.5}$. 

Other than the TESS data from Sectors 40, 47, 50 and 63 used in \cite{Maldonado2023}, we included the new light curves from Sectors 73 and 74 taken between UT 2023 December 7th and UT 2024 January 30$^{\rm th}$, all of which were taken with a 2 minute cadence and reduced by the Science Processing Operations Center \citep[SPOC;][]{Jenkins2016,Stumpe2012,Stumpe2014,Smith2012}. For consistency, we excluded the lower-cadence 30-minute data from Sectors 20 and 26. After masking all in-transit data, we fitted a Gaussian process (GP) model with the Mat\'{e}rn-3/2 kernel,  implemented with the \code{celerite} package \citep{Foreman2017} to detrend the simple aperture photometry of the TESS light curves without light dilution correction. We also made use of the publicly available ground-based photometric data taken in the Bessell I filter from the Red Buttes Observatory \citep[RBO;][]{Kasper2016}, published in \cite{Lambert2023}. For the spectroscopic data, except for the RM measurements obtained in this work (Section~\ref{spirou_data}), we incorporated the out-of-transit radial velocities (RVs) from the literature, including 12 HET/HPF observations from \cite{Lambert2023} as well as 9 TNG/HARPS-N measurements from \cite{Maldonado2023}. 

In the joint fit, we modeled 10 key parameters in total including orbital period ($P$), mid-transit epoch ($T_{0}$), radius ratio between the companion and the host star ($R_c/R_\ast$), the sum of host star and companion radii divided by the orbital semi-major axis ($(R_\ast+R_c)/a$), cosine of the orbital inclination ($\cos i_c$), RV semi-amplitude ($K$), two parameters related to eccentricity and argument of periastron ($\sqrt{e} \cos \omega$ and $\sqrt{e} \sin \omega$), the sky-projected spin-orbit angle ($\lambda$) and the projected stellar rotational velocity ($v\sin i$). For the photometry, we adopted a quadratic limb-darkening law for the TESS data \citep{Kipping2013} but a linear law for the ground-based RBO light curve, given its low-cadence and limited data points. Since the TESS simple aperture photometry has uncorrected light dilution effect due to the large pixel size ($21''$\,pixel$^{-1}$), we fitted a contaminating flux ratio $F_{C}/(F_C+F_T)$, where $F_T$ and $F_C$ are the flux of the target and nearby contaminating sources. The same parameter was fixed at zero instead for the RBO data since the target star was well resolved in the images. Moreover, we modeled the baseline offset as well as the jitter term for each photometric and spectroscopic dataset to take unaccounted white noise into consideration. Finally, we treated the SPIRou RM data taken on two nights as if they were from two different instruments since the SNRs and the observational conditions of two data sets are different. 

We placed wide uniform priors on all parameters, and determined the posteriors through a Markov Chain Monte Carlo (MCMC) analysis with the \code{emcee} package \citep{Foreman2013}. We initialized 60 walkers and each of them ran for 150,000 steps with the first 30,000 steps excluded as ``burn-in'' steps. The fit was considered as converged after all Markov chains were run for more than 30 times their autocorrelation length \citep{Foreman2013}. The spin-orbit angle was found to be $\lambda=-13.5^{+12.4}_{-13.8}\ ^{\circ}$, indicating an aligned geometry. The best-fit light contamination factor of the TESS photometry is $0.10\pm0.04$, consistent with the value 0.05 reported in the TESS input catalog \citep{Stassun2019tic}. We note that the RV jitter of the first SPIRou transit observation is much higher than the second visit (Figure~\ref{fig:joint-fit}), which might be partially explained by the persistence in the near infrared detector from another high SNR object observed before the first transit of TOI-5375. A tentative RV slope is present in the residuals of the first-night SPIRou data, although it is consistent with no slope at about the $1\sigma$ level. This residual slope probably originates from the systematic bias in the template-matching RV extraction methods \citep{Silva2025}, such as the LBL algorithm employed here \citep{Artigau2022}. Given the reasons outlined above, we emphasize that the first night SPIRou data should be used and interpreted with caution. We list the adopted priors and the results of key parameters in Table~\ref{allpriors}. Figure~\ref{fig:joint-fit} shows all datasets along with their best-fit models. 

\begin{table*}\scriptsize
    %\centering
    {\renewcommand{\arraystretch}{1.1}
    \caption{Parameter priors and best-fits in the joint model for \tar. $\mathcal{U}$(a, b) stands for a uniform prior between $a$ and $b$.}
    \begin{tabular}{lccr}
        \hline\hline
        Parameter       &Prior &Best-fit    &Description\\\hline
        \it{Host star parameters$^{[1]}$}\\
        $M_\ast$ ($M_\odot$) &$\cdots$ &$0.620\pm0.016$ &Stellar mass\\
        $R_\ast$ ($R_\odot$) &$\cdots$ &$0.649\pm0.024$ &Stellar radius\\
        $T_{\rm eff}$ ($K$) &$\cdots$ &$3897\pm88$ &Stellar effective temperature\\
        %$\log g_\ast$ (cgs) &$\cdots$ &$4.68\pm0.05$ &Stellar surface gravity\\
        %$P_{\rm rot}$ (days) &$\cdots$ &$1.972\pm0.008$ &Stellar rotation period \\
        %Age (Myr) &$\cdots$ &$\sim 400^{[2]}$ &Stellar age \\\hline
        \it{Key fitted parameters}\\
        $P$ (days)  &$\mathcal{U}$ ($1.0$\ ,\ $2.0$)  &$1.7215525_{-0.0000015}^{+0.0000017}$
        &Orbital period\\
        $T_{0}$ (BJD-2457000) &$\mathcal{U}$ ($2580.7$\ ,\ $2580.8$) &$2580.73654_{-0.00011}^{+0.00012}$ &Mid-Transit time\\
        %$r_1$ &$\mathcal{U}$ ($0$\ ,\ $1$) &$0.6662\pm0.0109$ &Parametrization for p and b.\\
        %$r_2$ &$\mathcal{U}$ ($0$\ ,\ $1$) &$0.1955\pm0.0008$ &Parametrization for p and b.\\
        $R_c/R_\ast$ &$\mathcal{U}$ ($0.0$\ ,\ $0.5$)  &$0.1812_{-0.0036}^{+0.0035}$
        &Companion-to-star radius ratio\\
        $(R_c+R_\ast)/a$ &$\mathcal{U}$ ($0.0$\ ,\ $0.5$)  &$0.1435_{-0.0037}^{+0.0034}$
        &Sum of radii divided by the orbital semi-major axis\\
        $\cos i_c$ &$\mathcal{U}$ ($0.0$\ ,\ $1.0$)  &$0.0531_{-0.0085}^{+0.0067}$
        &Cosine of the orbital inclination\\
        $\sqrt{e}\cos \omega$ &$\mathcal{U}$ ($-1$\ ,\ $1$) &$0.068_{-0.063}^{+0.026}$ &Parametrization for $e$ and $\omega$\\
        $\sqrt{e}\sin \omega$ &$\mathcal{U}$ ($-1$\ ,\ $1$) &$0.022_{-0.042}^{+0.039}$ &Parametrization for $e$ and $\omega$\\
        $K$ (m~s$^{-1}$) &$\mathcal{U}$ ($0$\ ,\ $30000$) &$18213_{-136}^{+104}$ &RV semi-amplitude\\
        $v\sin i$ (km s$^{-1}$) &$\mathcal{U}$ ($1$\ ,\ $100$) &$12.38_{-1.85}^{+2.05}$ &Projected stellar rotation velocity\\
        $\lambda$ (deg) &$\mathcal{U}$ ($-180$\ ,\ $180$) &$-13.5_{-13.8}^{+12.4}$ &Projected spin-orbit angle\\\hline
        %$v\sin i$ (km s$^{-1}$) &$\mathcal{U}$ ($0.1$\ ,\ $10$) &$1.65_{-0.09}^{+0.11}$ &Projected stellar rotation velocity\\\hline
        \it{Derived host star parameters}\\
        $i_\star$ (deg) &$\cdots$ &$53.4_{-11.3}^{+15.9}$ &Stellar inclination\\
        $\psi$ (deg) &$\cdots$ &$37.5_{-13.4}^{+10.6}$ &True obliquity\\\hline
        \it{Derived companion parameters}\\
        $R_{c}$ ($R_{J}$) &$\cdots$ &$1.17\pm0.05$ &Companion radius\\
        $M_{c}$ ($M_{J}$) &$\cdots$ &$84.8\pm1.5$ &Companion mass\\
        $q$ &$\cdots$ &$0.130\pm0.004$ &Companion-to-star mass ratio\\
        %$\rho_{p}$ (g~cm$^{-3}$) &$\cdots$ &$1.82\pm 0.19$ &Planet density.\\
        %$b$ &$\cdots$ &$0.499\pm0.016$ &Impact parameter.\\
        $a/R_\ast$ &$\cdots$ &$8.23\pm0.20$ &Scaled semi-major axis\\
        $a$ (AU) &$\cdots$ &$0.0250\pm0.0002$ &Semi-major axis\\
        $i_c$ (deg) &$\cdots$ &$87.0\pm0.5$ &Orbital inclination\\
        $e$ &$\cdots$ &$0.007\pm0.004$ &Orbital eccentricity\\
        %$\omega$ (rad) &$\cdots$ &$1.50_{-1.97}^{+0.06}$ &Argument of periapsis.\\
        %$T_{\rm eq}^{[3]}$ (K) &$\cdots$ &$728_{-44}^{+48}$ &Equilibrium temperature\\
        % %R_{p}/R_{\ast}$ &$\cdots$ &$0.1955\pm0.0008$ &Scaled planet radius.\\
        % $a/R_{\ast}$ &$\cdots$ &$13.694\pm0.151$ &Scaled semi-major axis.\\
        % $a$ (AU) &$\cdots$ &$0.040\pm 0.001$ &Semi-major axis.\\
        % $b$ &$\cdots$ &$0.499\pm0.016$ &Impact parameter.\\
        % $i$ (degrees) &$\cdots$ &$88.0\pm0.1$ &Orbital inclination.\\
        % $e$ &$\cdots$ &$0.041\pm0.015$ &Orbital eccentricity.\\
        % $\omega$ (degrees) &$\cdots$ &$130.0\pm44.5$ &Argument of periapsis.\\\hline
        % %\it{Planetary physical parameters}\\
        % $R_{p}$ ($R_{J}$) &$\cdots$ &$1.22\pm 0.04$ &Planet radius.\\
        % $M_{p}$ ($M_{J}$) &$\cdots$ &$2.48\pm 0.09$ &Planet Mass.\\
        % $\rho_{p}$ (g~cm$^{-3}$) &$\cdots$ &$1.82\pm 0.19$ &Planet density.\\
        % %$a$ (AU) &$\cdots$ &$0.040\pm 0.001$ &Semi-major axis.\\
        % $T_{\rm eq}^{[1]}$ (K) &$\cdots$ &$725\pm 20$ &Equilibrium temperature.\\
        % $S$ ($S_{\oplus}$) &$\cdots$ &$45.2\pm 3.2$ &Planetary Insolation.\\
        \hline%\hline
    \label{allpriors}    
    \end{tabular}}
    \begin{tablenotes}
       \item[1]  [1]\ The stellar parameters are adopted from \cite{Lambert2023}.  
       %\item[2]  [2]\ \cite{Maldonado2023} \todo{shows that \tar is not young}.
      %\item[3]  [3]\ We do not consider heat distribution between the dayside and nightside here and assume albedo $A_B=0$. 
    \end{tablenotes}
\end{table*}

\subsection{The Obliquity of the \tar\ System}

In the following analysis, we adopted the stellar parameters from \cite{Lambert2023} for simplicity but we note that the results are in accordance with those in \cite{Maldonado2023} within $1\sigma$. 

We first performed a generalized Lomb-Scargle (GLS) periodogram analysis \citep{Zechmeister2009} on all available 2-min TESS data. We derived a rotation period of $P_{\rm rot}=1.9691\pm0.0015$~days, in contrast to $1.9716^{+0.0080}_{-0.0083}$~days from \cite{Lambert2023} and $1.9692\pm0.0004$~days from \cite{Maldonado2023}. Combining the refined rotation period and the stellar radius $R_\ast=0.649\pm0.024\ R_\odot$ yields an equatorial rotation velocity of $v_{\rm eq}=2\pi R_{\ast}/P_{\rm rot}=16.7\pm0.7$~km s$^{-1}$, in agreement with the value $15.9\pm2.6$~km s$^{-1}$ reported by \cite{Maldonado2023}. In addition to the rotational velocity determined by the photometric modulation, \cite{Lambert2023} also measured the spectroscopic broadening velocity $v\sin i=16.7\pm0.9$~km s$^{-1}$ through empirical template match while we obtained a sky-projected stellar rotational velocity $v\sin i$ of $12.38^{+2.05}_{-1.85}$~km s$^{-1}$ based on our joint analysis, slightly smaller but consistent within about $2\sigma$.

Following the Bayesian methodology presented in \cite{Masuda2020}, we then determined the cosine of stellar inclination $\cos i_{\star}$ based on the known priors of $R_\ast$ from \cite{Lambert2023}, $P_{\rm rot}$ and $v\sin i$ from this work. We sample the posterior distribution using MCMC with wide, non-informative priors on all three input parameters. We obtained $\cos i_{\star}=0.60^{+0.15}_{-0.24}$, corresponding to a stellar inclination of $i_\star=53.4^{+15.9}_{-11.3}\ ^{\circ}$. The three-dimensional true obliquity $\psi$ was then calculated via the relation in \cite{Albrecht2022}:
\begin{equation}
    \cos (\psi) = \cos (i_{\star})\cos (i_{c})+\sin (i_{\star})\sin (i_{c})\cos (\lambda),
\end{equation}
together with the posteriors on the orbital inclination $i_{c}$ and the projected spin-orbit angle $\lambda$ from the joint fit. This yields a 3D true obliquity $\psi=37.5^{+10.6}_{-13.4}\ ^{\circ}$. Given the large uncertainty on both projected and deprojected obliquities dominated by the low precision of RM data due to the faintness of the host star, we thus conservatively report that the orbit is aligned. Future observations with larger telescopes or more visits from SPIRou are required to reach higher SNR and better constrain the stellar obliquity. 

% \begin{equation}
% \begin{aligned}
%     \mathcal{L} = & \left(\frac{R_\ast/R_{\odot}-0.649}{0.024}\right)^{2}+\left(\frac{P_{\rm rot}-1.9691\ {\rm days}}{0.0015\ {\rm days}}\right)^{2} \\
%     &+\left(\frac{v\sqrt{1-\cos^{2} i_\star}-12.38\ {\rm km/s}}{2.05\ {\rm km/s}}\right)^{2}.
% \end{aligned}
% \end{equation}

\begin{figure*}[htb]
    \centering
    \includegraphics[width=0.99\linewidth]{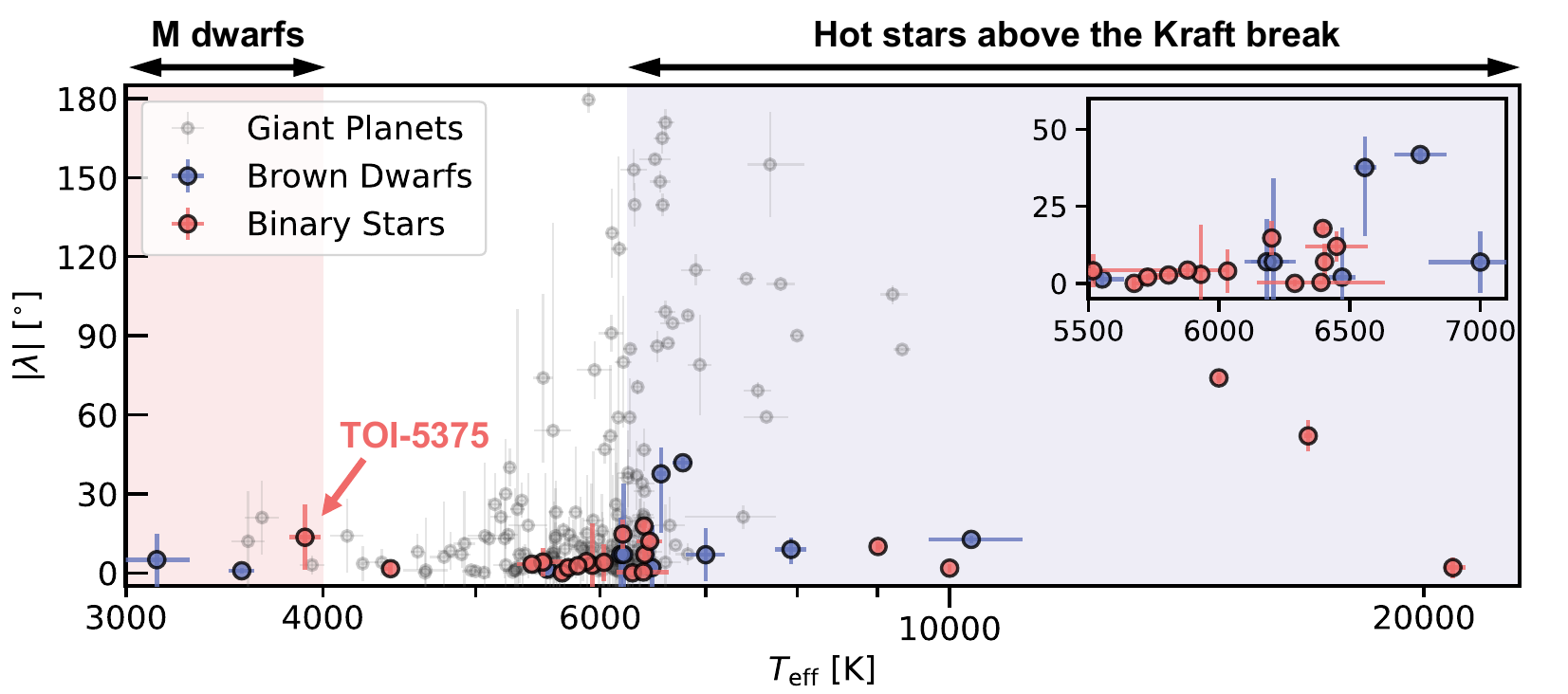}
    \caption{Projected stellar obliquity ($\lambda$) of giant-planet ($0.3\leq M_c< 13.6\,M_J$, gray), brown dwarf ($13.6\leq M_c< 80\,M_J$, blue) and binary star ($M_c\geq80\,M_J$, red) systems as a function of primary star effective temperature. M dwarfs and hot stars with effective temperatures above the Kraft break ($T_{\rm eff}\sim 6250\ K$) are shown as red and blue shaded regions, respectively. \tar\ is marked with a red arrow. The obliquities of giant planets and brown dwarfs are retrieved from the TEPCat catalog \cite{Southworth_2011} while the results of binary stars come from \cite{Marcussen2022} and some recent works \citep{Spejcher2025,Spejcher2025b,Wells2025}.}
    \label{fig:lambda_teff}
\end{figure*}

\begin{figure*}[htb]
    \centering
    \includegraphics[width=\linewidth]{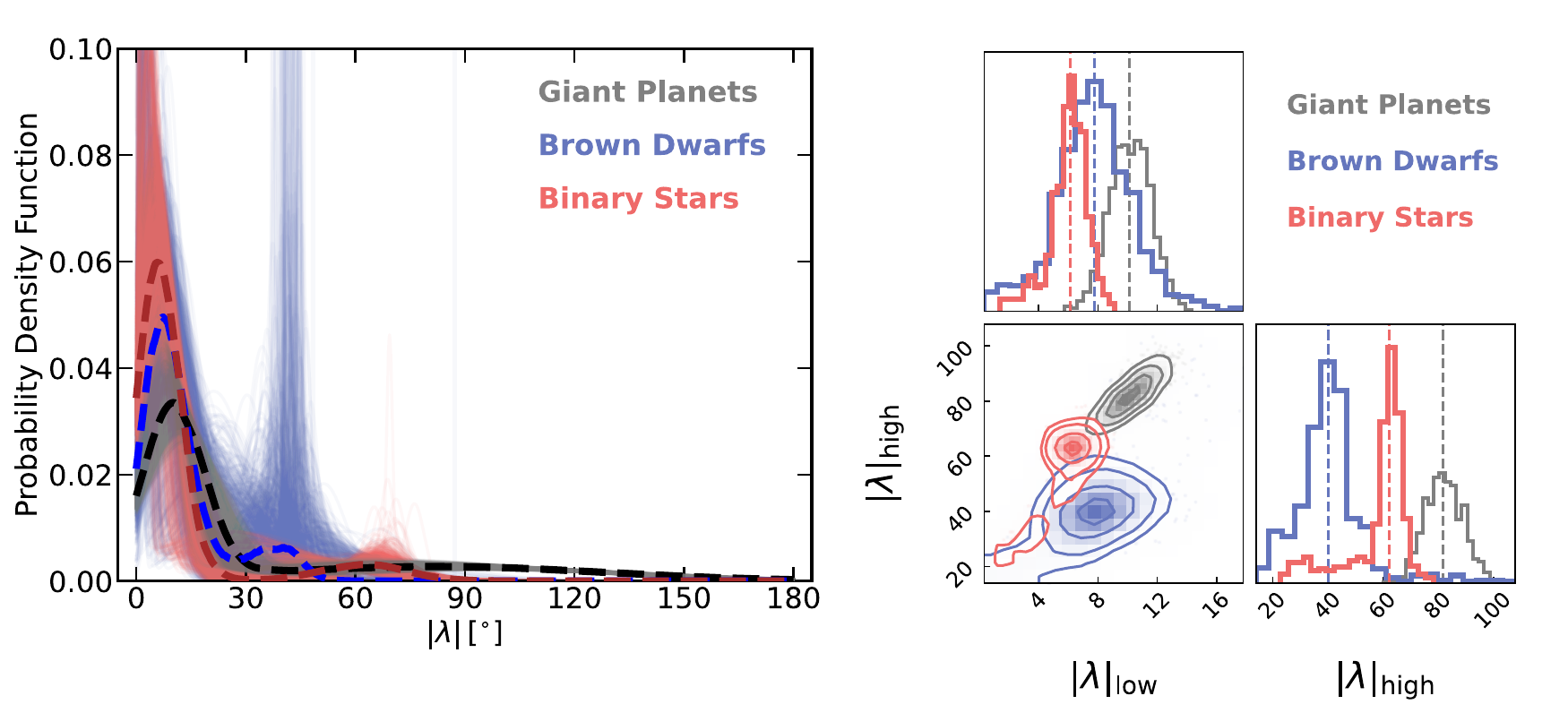}
    \caption{Left panel: Gaussian mixture model fitting to the $|\lambda|$ distribution of giant planets (gray, $0.3\leq M_c<13.6\,M_J$), brown dwarfs (blue, $13.6\leq M_c<80\,M_J$) and binary stars (red, $M_c\geq 80\,M_J$), based on 2000 randomly generated synthetic datasets (see Section~\ref{pattern} for details). The median results are shown as dashed curves. The low-$|\lambda|$ components of binary stars and brown dwarfs are slightly more concentrated and close to zero compared with the giant planet population. Right panel: The low and high obliquity component's mean value distributions of giant planets (gray), brown dwarfs (blue) and binary stars (red), based on two-component Gaussian mixture model fittings to 1000 simulated synthetic datasets. The vertical dashed lines show the 50th quantiles of the distributions.}
    \label{fig:gaussian_mixture}
\end{figure*}

\begin{figure*}[htb]
    \centering
    \includegraphics[width=0.99\linewidth]{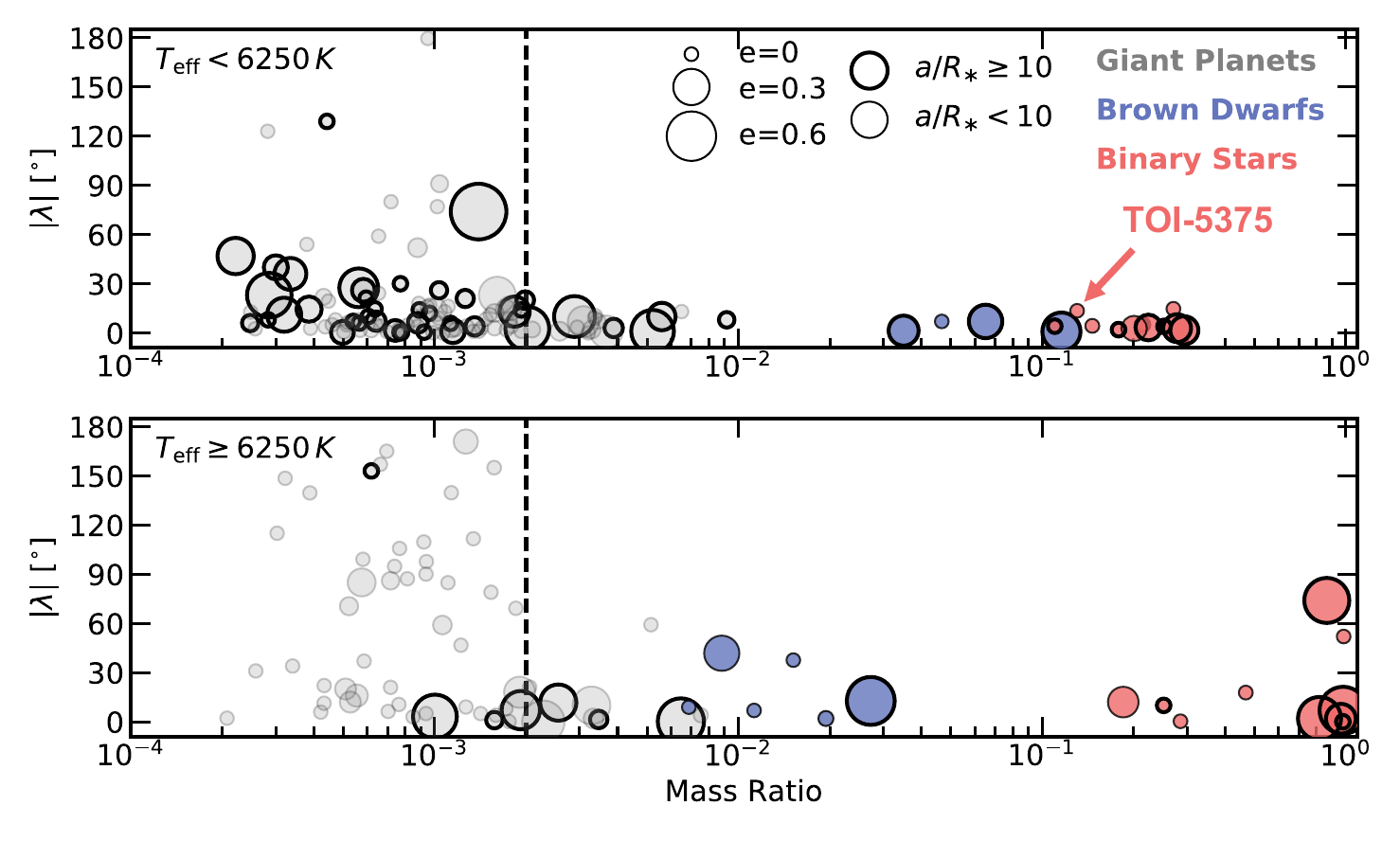}
    \caption{Projected stellar obliquity ($\lambda$) of three companion classes as a function of the companion-to-primary mass ratio. The top and bottom panels shows the systems with cool and hot primary stars, defined according to the Kraft break. The size of the symbol is proportional to the orbital eccentricity while the dots highlighted by a thick boundary represent companions with large scaled semi-major axis ($a/R_\ast\geq10$). The vertical black dashed line marks the empirical boundary at $2\times 10^{3}$ that separates aligned high-mass-ratio systems and misaligned low-mass-ratio systems, proposed by \cite{Rusznak2025} based on a mixed giant planet and brown dwarf sample. The low-mass-ratio systems exhibit more dispersed obliquities regardless of stellar temperature. Wide-orbit companions, regardless of their masses, are likely aligned. No significant correlation between eccentricity and obliquity is seen in the sample.}
    \label{fig:lambda_q}
\end{figure*}

The joint fit reveals that the companion has a radius of $1.17\pm0.05\ R_J$ and a mass of $84.8\pm1.5\ M_J$ on a circular orbit, leading to a mass ratio of $0.130\pm0.004$. We designate the companion as a very-low-mass M star for two reasons. First, the companion mass exceeds the hydrogen fusion limit of approximately $80\ M_J$ \citep{Laughlin1997}. Based on the joint-fit results, we further carried out a secondary eclipse analysis. We measured a significant secondary eclipse signal with a depth of $1756\pm135$ ppm, after accounting for the light dilution effect. Assuming blackbody spectra without reflected starlight, such a signal in the TESS spectral bandpass is consistent with a stellar companion having an effective temperature of $2370\pm 100\ K$ \citep{Charbonneau2005}, corresponding to an M9V main sequence star \citep{Pecaut2013}. Figure~\ref{fig:lambda_teff} shows the projected stellar obliquity $\lambda$ and host star effective temperature diagram of all giant planets, brown dwarfs, and binary stars. Prior to this work, obliquity studies existed for three hot Jupiters (TOI-4201, \citealt{Gan2024}; TOI-3714 and TOI-5293A, \citealt{Weisserman2025}) and two brown dwarfs (LP~261-75, \citealt{Brady2025} and TOI-2119, \citealt{Doyle2025}) orbiting M stars, all of which showed well-aligned orbits. With the measurement reported here, \tar\ represents the first M-dwarf binary system with a determined stellar obliquity.

The good alignment ($\lambda=-13.5_{-13.8}^{+12.4}\,^{\circ}$ and $\psi=37.5^{+10.6}_{-13.4}\,^{\circ}$) of the \tar\ system suggests 
two possible evolutionary histories: a dynamically quiescent formation with primordial alignment, or subsequent tidal realignment. Following the equation derived in \cite{Zahn1977}, we estimate that the tidal realignment timescale $\tau_{\rm realign}$ of the \tar\ system is $46\pm11$ Myr. Although the short rotation period of \tar\ is suggestive of a young host M dwarf with an age of about 400 Myr \citep{Lambert2023}, \cite{Maldonado2023} argued that \tar\ appears to be an old field star according to its stellar metal content and kinematics information. Regardless of this age discrepancy, the obliquity damping timescale is significantly smaller than the age, indicating that the system has sufficient time to realign.

\section{Obliquity Patterns Across Giant Planets, Brown Dwarfs and Binary Stars}\label{pattern}

In this section, we compare the projected stellar obliquities of three populations categorized by companion mass $M_c$: giant planets ($0.3\leq M_{c}<13.6\ M_J$), brown dwarfs ($13.6\leq M_{c}<80\ M_J$) and binary stars ($M_{c}\geq80\ M_J$). Our obliquity sample for giant planets and brown dwarfs is retrieved from the SOCat\footnote{\url{https://www.stellarobliquity.com/}}, a stellar obliquity catalog based on the TEPCat catalog\footnote{\url{https://www.astro.keele.ac.uk/jkt/tepcat/}} \citep{Southworth_2011}. The binary star sample is compiled from \cite{Marcussen2022} supplemented by several recently published works \citep{Spejcher2025,Spejcher2025b,Wells2025}.  If multiple obliquity measurements of a single system were available, we selected the results from RM first, followed by Doppler tomography and other methods. We further excluded the systems that only have upper limit constraints on companion mass or orbital eccentricity. The final sample comprises 173 giant planets, 11 brown dwarfs and 21 binary stars. For binary systems, we consider only the obliquity of the primary star. Throughout the work, we define a misaligned system if its absolute sky-projected obliquity is above 30 degrees (i.e., $|\lambda|\geq30^{\circ}$). 

\subsection{Obliquity vs. Effective Temperature}

One of the most important findings in stellar obliquity studies is the dependence on effective temperature. Hot Jupiters orbiting stars with effective temperature above the Kraft break \citep[$T_{\rm eff} \gtrsim 6250$\,K;][]{Kraft1967} exhibit a broad range of spin–orbit angles \citep{Winn2010,Albrecht2012,Wang2026}, while the counterparts around cool stars mostly have aligned orbits. Such a characteristic might be due to the differences in stellar interior structures. Cool stars, especially M dwarfs, possess deep convective zones \citep{Pinsonneault2001} that enable rapid tidal dissipation. Therefore, the stellar obliquities could be efficiently erased even if the hot Jupiters around cool stars were born with misalignments \citep{Albrecht2012,Wang2021K2_232}. On the other hand, internal gravity waves (IGWs) probably also play a role under this framework. Hot stars have convective cores surrounded by extended radiative envelopes outside, where angular momentum transport via IGWs at the convective–radiative boundary can excite the spin of the host star \citep{Rogers2012,Rogers2013b}. An alterative hypothesis, resonance locking, has also been proposed to explain the $\lambda$-$T_{\rm eff}$ relationship: by increasing their stellar gravity mode frequency, cool stars eventually undergo strong tidal evolution, dampening the obliquity \citep{Zanazzi2024}.

Brown dwarfs and binary stars probably exhibit a similar paradigm. As shown in Figure~\ref{fig:lambda_teff}, moderate misalignments with $|\lambda|\geq 30 ^{\circ}$ have been detected only for brown dwarfs (CoRoT-3, \citealt{Triaud2009} and XO-3, \citealt{Winn2009,Hirano2011,Rusznak2025}) and binaries (DI Herculis, \citealt{Albrecht2009} and CV Velorum, \citealt{Albrecht2014}) around hot stars. However, we caution the readers that only two brown dwarfs and two binaries have misaligned orbits, hence the sample is too limited to make any robust conclusions so far. Given the possible distinct $|\lambda|$ behaviors below and above the Kraft break across all three populations, we attempted to fit two-component Gaussian mixture models to the absolute sky-projected obliquity distributions of each companion class using the \code{GaussianMixture} function embedded in \code{scikit-learn} \citep{scikit-learn}. To account for the uncertainty of each $\lambda$ measurement, we employed a Monte Carlo resampling approach. For every system, we randomly generate 1000 synthetic results based on a Gaussian distribution $\mathcal{N}(\lambda,\sigma_{\lambda}^{2})$, centering at the reported measurements with $\sigma_{\lambda}$ taken as the larger value of the upper and lower errors. We looped the fitting 1000 times and recorded the resulting models. The left panel of Figure~\ref{fig:gaussian_mixture} shows the best-fit probability density functions of $|\lambda|$ for the three companion groups while the distributions of mean of each mixture component are presented in the right. Compared to brown dwarfs and giant planets, the low-$|\lambda|$ component for binary stars is more concentrated and close to zero, with peak values of $|\lambda|_{\rm low}$ at $6.3\pm1.2^{\circ}$, $8.0\pm3.1^{\circ}$, and $10.1\pm1.4^{\circ}$ for binary stars, brown dwarfs and giant planets, respectively. This suggests that some binary stars and brown dwarfs were probably born more aligned than giant planets, or they went through more efficient tidal realignment due to the shorter timescale (see below). Meanwhile, the high-$|\lambda|$ peaks for brown dwarfs and binary stars are not yet statistically robust due to the limited sample size, particularly of misaligned systems. In contrast, the high-$|\lambda|$ component for giant planets is broad, with a tentative peak of $|\lambda|_{\rm high}$ at $81.8\pm 6.8^{\circ}$, though the result is subject to the choice of priors and models. Several studies have examined the stellar obliquity distribution, particularly the deprojected obliquity $\psi$, of planetary systems in detail, finding no evidence for a perpendicular peak near $90^{\circ}$ through hierarchical Bayesian analysis \citep[e.g.,][]{Dong2023,Siegel2023}. With a desired larger sample in the future, a similar investigation into the stellar obliquity distribution of brown dwarf and binary star systems would be valuable. Nonetheless, there is one key qualitative difference: retrograde orbits with $|\lambda|\geq 90 ^{\circ}$ are observed among giant planet systems but rarely seen in sub-stellar systems and stellar binaries. 

\subsection{Obliquity vs. Mass Ratio}

Planetary systems with high mass ratios ($q>2\times 10^{-3}$) tend to show low stellar obliquities \citep[e.g.,][]{Hebrard2011_HATP6,Gan2024,Rusznak2025}, regardless of the host star's effective temperature. For stars with convective envelopes, this could be a natural consequence of tidal damping, as the realignment timescale scales with $\tau_{\rm realign}\propto q^{-2}$ \citep{Zahn1977}, leading to faster realignment for higher-mass-ratio systems. Alternatively, it may reflect a primordial difference: lower-mass-ratio systems could have experienced dynamical instabilities in compact configurations, while higher-mass-ratio systems formed in isolation and retained their initial alignment, as suggested by \cite{Rusznak2025}. Figure~\ref{fig:lambda_q} presents the projected obliquity and mass ratio distributions of three companion groups. All brown dwarf and stellar binary systems lie above $2\times 10^{-3}$, and they appear to be more aligned than systems with mass ratios below this empirical boundary, all of which are giant planets. Notably, the subset of systems orbiting cool host stars ($T_{\rm eff} < 6250$ K) exhibits a tighter, less dispersed obliquity distribution.

\subsection{Obliquity vs. Eccentricity and Semi-Major Axis}

As mentioned above, giant planets encounter dynamical instabilities that may lead to misaligned orbits. One might expect a correlation between obliquity and orbital eccentricity, which could also be excited through processes like scattering \citep{Rasio1996,Chatterjee2008,Ford2008}, Kozai–Lidov interactions \citep{Fabrycky2007,Naoz2016} and secular resonances \citep{Wu2011,Petrovich2020}. We therefore investigate whether a $|\lambda|$–$e$ correlation exists across the three companion populations. In Figure~\ref{fig:lambda_q}, we further illustrate the eccentricity of each system using different symbol size. We find no significant dependence in any of three groups, although a recent study focusing on astrometric binaries shows that misalignments are more frequent if the orbits have higher eccentricities \citep{Marcussen2024}. Specifically, only six giant planets, one brown dwarf and one stellar binary systems have projected obliquities $|\lambda|\geq30^{\circ}$ while having robust eccentric orbits with $e>0.1$. However, this finding is probably due to observational bias. Obliquity measurements are biased toward short-period systems, which are subject to the tidal circularization effect \citep{hut81,Jackson2008}. Since the orbital scaled semi-major axis ($a/R_\ast$) is a fundamental parameter relevant to the tidal circularization timescale, we separate the sample into two subgroups based on a cut at $a/R_\ast=10$. We examine the obliquities of long-period systems with $a/R_\ast\geq 10$ (highlighted dots in Figure~\ref{fig:lambda_q}), which are expected to preserve the initial dynamical state of orbital eccentricity and misalignment. Previous works have pointed out that warm Jupiters tend to be more aligned than hot Jupiters \citep{Rice2022warmjupiter,Wang2024}, indicating that giant planets are probably formed within aligned protoplanetary disks. Although the formation channels are different as giant planets, the vast majority of long-period brown dwarfs and binary stars are likewise well-aligned \citep[see also][]{Vowell2026}, suggesting a similar history of formation within aligned disks. Among a total of 5 brown dwarfs and 12 binary stars with $a/R_\ast\geq 10$, a notable exception is the highly eccentric and misaligned DI Herculis binary system, which likely owes its architecture to perturbations from a tertiary companion \citep{Albrecht2009}. Indeed, the triple/high-order fraction for a massive star like DI Herculis is about 40\% \citep{Offner2023}. Nevertheless, given the small sample of long-period brown dwarfs and binary stars, we are not able to draw a firm conclusion at this point. Extending the obliquity study to cover more such systems shall remedy the situation.

%some words about realignment and IGW. tidal realignment. orbital excitation through a third body.
%\todo{Doug suggests two papers \cite{Rogers2012,Rogers2013a,Rogers2013b}:internal gravity waves can explain the misalignment of hot Jupiters around hot stars, especially those with retrograde orbits.} This mechanism has nothing to do with the companion and the orbit. it only excite the spin of the host star. Only hot stars works because they have radiative zones outside the convective zones. Very hard to see around M dwarfs because they are fully convective.

% \todo{5 key points I should cover:
% 1.Giant planets, brown dwarfs and binary stars tend to have aligned orbits when the primary stars are cool
% 2. xxxx when the mass ratios are high 
% 3. Retrograde orbits are mostly seen in giant planet systems.
% 4. No clear correlation between projected obliquity and eccentricity
% in all three categories
% 5. Long-period companions seem to be more aligned.}

\section{Conclusions}\label{conclusion}

In this work, we present the first Rossiter–McLaughlin measurement of an M dwarf binary system. Based on CFHT/SPIRou spectroscopic observations, we found that the system, \tar, has a projected obliquity of $\lambda=-13.5_{-13.8}^{+12.4}\,^{\circ}$. Together with the information of stellar rotation, we further determined a true 3D obliquity of $\psi=37.5^{+10.6}_{-13.4}\,^{\circ}$. Both results indicate an aligned orbit, implying that the companion \tar B either formed with a primordially aligned configuration or underwent tidal realignment during its evolution.

Building on the result of \tar, we conducted a comparative analysis of stellar obliquities across giant planets, brown dwarfs as well as stellar binaries. We summarize the key features below:

\begin{enumerate}[(i)]
\item Similar to giant planets, brown dwarfs and binary stars orbiting cool stars with $T_{\rm eff}<6250\,K$ are probably predominantly aligned but a larger sample is required to reach a robust conclusion.
\item Assuming the underlying $|\lambda|$ distribution can be described by a two-component Gaussian mixture model, the low-$|\lambda|$ components for binary stars and brown dwarfs are more concentrated and close to zero than that of giant planets.
%\item Retrograde orbits have only been detected in giant planet systems.
\item High-mass-ratio systems with $q > 2 \times 10^{-3}$, encompassing all brown dwarfs and stellar binaries in our sample, exhibit a stronger tendency toward alignment compared to lower-mass-ratio giant planets.
\item No clear correlation between projected obliquity and eccentricity is seen in any of the three populations.
\item The vast majority of long-period companions with scaled semi-major axis $a/R_\ast\geq10$ are aligned, a trend that appears independent of companion mass.
\end{enumerate}

%% Also note that the akcnowlodgment environment does not support long amounts of text. If you have a lot of people and institutions to acknowledge, do not use this command. Instead, create a new \section{Acknowledgments}.

\section{Acknowledgments}

We thank Doug Lin, Sergei Nayakshin, Xianyu Wang and Mika Lambert for the useful discussions.

T.G. and S.M. acknowledge support by the National Natural Science Foundation of China (No. 12133005). A.L., É.A., C.C., R.D. and N.J.C. acknowledge the financial support of the Fonds de recherche du Québec - Secteur Nature et technologies (FRQ-NT) through the Centre de recherche en astrophysique du Québec as well as the support from the Trottier Family Foundation and the Trottier Institute for Research on Exoplanets. A.L. acknowledges support from the FRQ-NT under file \#349961. É.A. and R.D. acknowledge support from Canada Foundation for Innovation (CFI) program, the Université de Montréal and Université Laval, the Canada Economic Development (CED) program and the Ministere of Economy, Innovation and Energy (MEIE).

%SPIRou observations:
Based on observations obtained at the Canada-France-Hawaii Telescope (CFHT) which is operated from the summit of Maunakea by the National Research Council of Canada, the Institut National des Sciences de l'Univers of the Centre National de la Recherche Scientifique of France, and the University of Hawaii. The observations at the Canada-France-Hawaii Telescope were performed with care and respect from the summit of Maunakea which is a significant cultural and historic site. Based on observations obtained with SPIRou, an international project led by Institut de Recherche en Astrophysique et Planétologie, Toulouse, France.

% ExoFOP, % MAST
Funding for the TESS mission is provided by NASA's Science Mission Directorate. We acknowledge the use of public TESS data from pipelines at the TESS Science Office and at the TESS Science Processing Operations Center. Resources supporting this work were provided by the NASA High-End Computing (HEC) Program through the NASA Advanced Supercomputing (NAS) Division at Ames Research Center for the production of the SPOC data products. This research has made use of the Exoplanet Follow-up Observation Program website, which is operated by the California Institute of Technology, under contract with the National Aeronautics and Space Administration under the Exoplanet Exploration Program. This paper includes data collected by the \tess\ mission, which are publicly available from the Mikulski Archive for Space Telescopes\ (MAST).

% %ExoFOP
% Funding for the TESS mission is provided by NASA's Science Mission Directorate. This research has made use of the Exoplanet Follow-up Observation Program website, which is operated by the California Institute of Technology, under contract with the National Aeronautics and Space Administration under the Exoplanet Exploration Program. This paper includes data collected by the \tess\ mission, which are publicly available from the Mikulski Archive for Space Telescopes\ (MAST). 

%\todo{To be filled in.}


\vspace{5mm}
\facilities{CFHT/SPIRou, HET/HPF, TNG/HARPS-N, TESS, RBO}


\software{Allesfitter \citep{Gunther2021}, corner \citep{Foreman2016}, emcee \citep{Foreman2013}, ellc \citep{Maxted2016}, scikit-learn \citep{scikit-learn}
          }



%\clearpage

\appendix

\counterwithin{table}{section}
\counterwithin{figure}{section}

\section{Vetted Template Construction}
\label{appendix:vetted_template}

The LBL radial velocity extraction method requires a high-SNR stellar template that accurately represents the intrinsic spectrum of the target star with pixel-to-pixel noise significantly smaller than the flux gradient for typical spectral features. For faint or rapidly rotating stars---where line depths are reduced---constructing such a template directly from the observations can be challenging. We describe a method for constructing optimized LBL  templates  when the dataset in hand is insufficiently large to obtain a  high enough SNR through a median combination of all spectra in hand.

\subsection{Method Overview}

We model the target spectrum as a  combination of high-SNR templates from reference stars with low $v \sin i$, convolved with a rotational broadening kernel and a two-component telluric absorption model. The reference library comprises 15 high-SNR stellar templates constructed from archival SPIRou spectra, registered to a common systemic velocity, spanning spectral types K2V to M7V. For a late-M dwarf, the earliest-type reference (K2V) will receive a vanishingly small weight in the linear combination, while later-type templates dominate.

Spectra are processed in logarithmic flux space following high-pass filtering to remove the continuum. Computations are performed on  pre-telluric-correction spectra; this will facilitate future integration of vetted templates within the \code{APERO} framework \citep{Cook2022} to improve telluric absorption and sky emission correction. The vetted template methodology is fully operational within the LBL framework, its implementation within \code{APERO} remains under development.\footnote{The vetted template code is publicly available at \url{https://github.com/eartigau/vettedtemplate}.}

\subsection{Template optimisation}

For each observed spectrum, we  determine the systemic radial velocity through a grid search. The reference templates are Doppler-shifted across a range of velocities  and vsini values (\citet{Gray2005} formalism), and at each step perform  a  least-squares fit to the optimal combination of template and telluric components. 

\begin{figure}[!htbp]
    \centering
    \includegraphics[width=\linewidth]{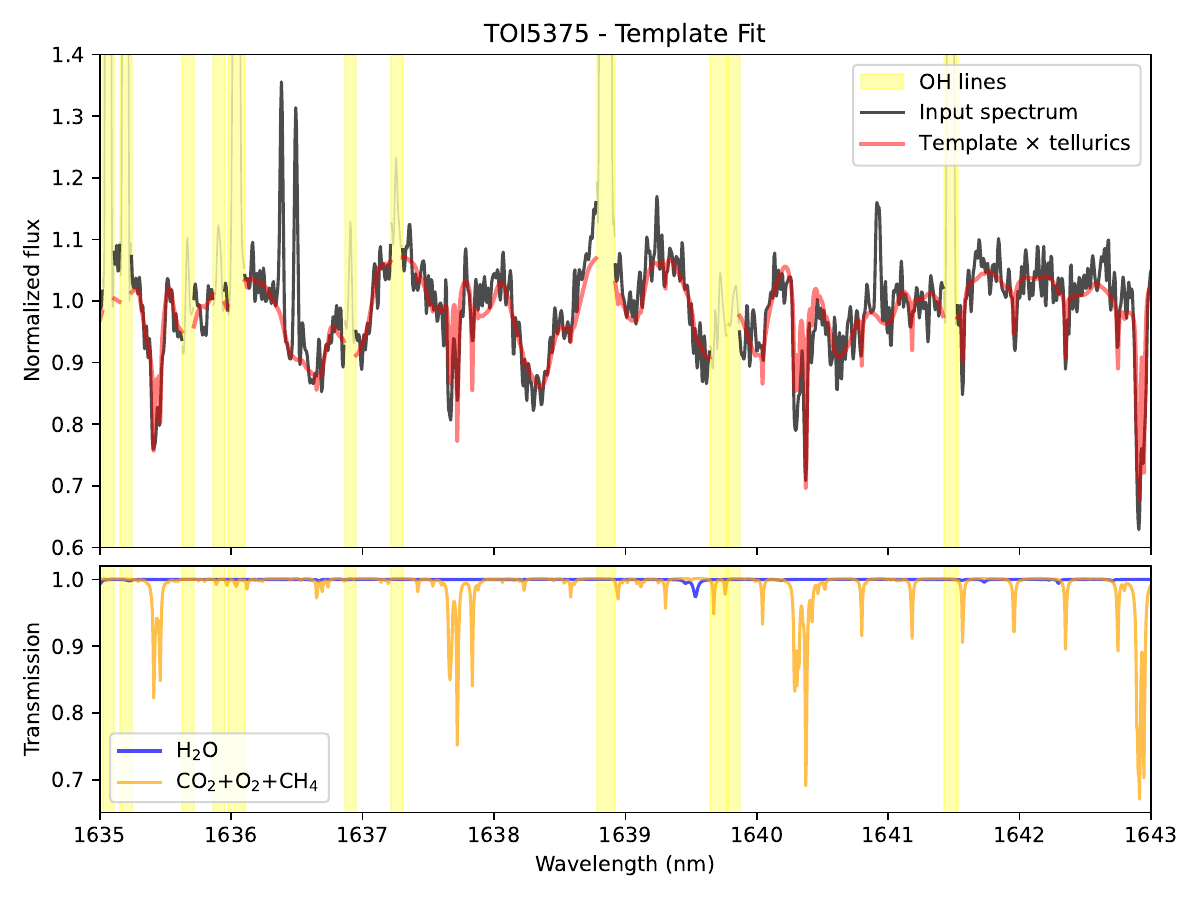}
    \caption{Template for a representative observation of TOI-5375. The observed spectrum (black) is well reproduced by the model combining reference templates and telluric absorption (red). Sky emission lines are present in the domain; we reject around lines listed in \citet{Rousselot2000} (yellow shading). A handful of emission lines are present and unaccounted for, but these have basically no impact on the robust fitting.}
    \label{fig:template_fit}
\end{figure}

At each step, the observed spectrum is fitted via bounded linear least-squares optimization of Doppler-shifted and rotationally broadened reference templates, water vapor, and dry absorbers (CO$_2$ + O$_2$ + CH$_4$) from the TAPAS atmospheric model \citep{Bertaux2014}:
\begin{equation}
    \log F_\mathrm{obs}(\lambda) = \sum_{i} a_i \, \tau_i + \mathrm{continuum},
\end{equation}
where $a_i \geq 0$ are bound to be positive. The optimum is found when the robust standard deviation\footnote{Defined as half of the distance between the 16$^{\rm th}$ and 84$^{\rm th}$ percentile of the distribution} of model-to-template residual is minimized. Figure~\ref{fig:template_fit} shows an example fit for TOI-5375.

\subsection{Template Construction}

Following the analysis of all available observations of the target, the individual spectral fits are combined to produce the final template. Each observation is weighted according to its fit quality (inverse-RMS weighting), and outlier spectra are rejected via sigma-clipping of the derived parameters. The resulting template inherits the high SNR of the reference library while being tailored to the specific spectral characteristics of the target star. This vetted template subsequently serves as input to the LBL radial velocity extraction framework.

In addition to providing a high-SNR template, the procedure also determines a mean systemic velocity and $v\sin i$ with corresponding dispersions. The values derived for TOI-5375 are $v_{\rm sys} = -63.5\pm4.0$\,km\,s$^{-1}$ and $v\sin i = 17.66\pm0.46$\,km\,s$^{-1}$, consistent with $v\sin i=16.7\pm0.9$~km\,s$^{-1}$ measured by \citet{Lambert2023}.

\section{SPIRou RVs and Activity Indicators}
\label{appendix:RVs}

Table~\ref{RVtable} lists the SPIRou RVs and stellar activity indicators of \tar. 

\begin{table*}[htb]
    %\centering
    {\renewcommand{\arraystretch}{0.85}
    \caption{SPIRou radial velocities and stellar activity indicators.}
    \begin{tabular}{ccccccc}
        \hline\hline
        BJD       &RV (m~s$^{-1}$) &$\sigma_{\rm RV}$ (m~s$^{-1}$) &CRX &$\sigma_{\rm CRX}$ &dTemp &$\sigma_{\rm dTemp}$ \\\hline
        UT 2025 February 10th & &\\
2460716.863249 	&-57547.64 	&59.32 	&-824.29 	&448.72 	&5.44 	&2.92 \\
2460716.873957 	&-57969.18 	&57.84 	&191.92 	&430.86 	&2.52 	&2.84 \\
2460716.884729 	&-58771.14 	&58.11 	&-133.03 	&442.99 	&3.05 	&2.85 \\
2460716.895437 	&-59437.58 	&59.65 	&-206.79 	&454.89 	&8.44 	&2.92 \\
2460716.906145 	&-60290.79 	&78.77 	&-987.83 	&644.13 	&2.01 	&3.83 \\
2460716.916851 	&-60659.22 	&63.84 	&-1410.95 	&486.53 	&5.14 	&3.12 \\
2460716.927559 	&-61618.93 	&58.05 	&-1248.29 	&442.53 	&4.75 	&2.84 \\
2460716.938268 	&-61932.69 	&60.79 	&-914.53 	&458.85 	&5.66 	&2.97 \\
2460716.948974 	&-62558.02 	&62.62 	&-910.82 	&474.07 	&-5.70 	&3.07 \\
2460716.959681 	&-63501.89 	&63.51 	&-1285.43 	&478.35 	&-3.66 	&3.10 \\
2460716.970389 	&-64474.35 	&62.71 	&-411.48 	&462.77 	&-3.23 	&3.08 \\
2460716.981100 	&-64980.78 	&61.87 	&-694.48 	&452.60 	&2.35 	&3.01 \\
2460716.991803 	&-65800.42 	&60.35 	&-650.36 	&439.19 	&5.93 	&2.96 \\
2460717.002515 	&-66291.55 	&66.00 	&-566.00 	&483.38 	&7.46 	&3.25 \\
2460717.013222 	&-67255.69 	&64.39 	&-1019.51 	&483.09 	&7.41 	&3.13 \\
2460717.023929 	&-67876.30 	&66.18 	&-384.79 	&495.01 	&8.28 	&3.26 \\
2460717.034637 	&-68240.05 	&68.81 	&-933.69 	&526.61 	&4.13 	&3.36 \\
2460717.045344 	&-68871.91 	&67.59 	&-941.31 	&525.87 	&9.77 	&3.31 \\
2460717.056051 	&-69724.91 	&71.76 	&730.43 	&552.27 	&5.50 	&3.55 \\\hline
        UT 2025 February 17th & &\\
2460723.760262 	&-57694.58 	&57.46 	&-312.49 	&388.76 	&14.73 	&2.86 \\
2460723.770973 	&-58476.18 	&57.86 	&-978.67 	&395.20 	&9.42 	&2.89 \\
2460723.781676 	&-59280.35 	&59.87 	&-661.49 	&415.73 	&9.28 	&2.98 \\
2460723.792384 	&-60033.25 	&57.52 	&-1249.13 	&389.78 	&7.08 	&2.83 \\
2460723.803091 	&-60694.70 	&57.34 	&-233.57 	&389.22 	&6.32 	&2.84 \\
2460723.813798 	&-61223.81 	&56.96 	&-866.97 	&390.18 	&7.35 	&2.86 \\
2460723.824506 	&-61810.92 	&56.89 	&-1109.35 	&389.56 	&3.33 	&2.83 \\
2460723.835213 	&-62508.05 	&57.95 	&-527.84 	&403.13 	&4.61 	&2.88 \\
2460723.845926 	&-63361.17 	&56.72 	&-1438.12 	&390.73 	&-1.57 	&2.81 \\
2460723.856628 	&-64376.82 	&59.94 	&-930.38 	&428.56 	&-5.56 	&2.95 \\
2460723.867334 	&-65073.35 	&56.29 	&-1214.98 	&398.70 	&-1.16 	&2.78 \\
2460723.878043 	&-65733.04 	&53.37 	&-141.05 	&387.87 	&7.69 	&2.65 \\
2460723.888750 	&-66331.45 	&54.21 	&-336.97 	&399.41 	&2.24 	&2.69 \\
2460723.899458 	&-67052.93 	&53.60 	&-1422.61 	&393.24 	&7.95 	&2.68 \\
2460723.910165 	&-67730.37 	&52.97 	&-430.95 	&381.21 	&3.76 	&2.65 \\
2460723.920874 	&-68443.90 	&55.21 	&-1028.28 	&399.53 	&5.86 	&2.73 \\
2460723.937707 	&-69439.86 	&55.09 	&-108.55 	&397.01 	&3.71 	&2.74 \\
2460723.948414 	&-70179.29 	&57.65 	&-642.01 	&427.45 	&4.61 	&2.87 \\
2460723.959123 	&-70858.40 	&54.80 	&-681.52 	&398.96 	&9.27 	&2.74 \\
        
         \hline%\hline 
    \label{RVtable}
    \end{tabular}}
\end{table*}

\bibliography{planet}{}
\bibliographystyle{aasjournal}



\end{document}